\documentclass[showpacs,amsmath,amssymb,aps,twocolumn,superscriptaddress,prx]{revtex4-2}
\usepackage{mathtools}
\usepackage{algorithm}
\usepackage{algpseudocode}
\usepackage{multirow}
\usepackage[pdftex]{graphicx} \graphicspath{{}}
\usepackage{grffile} 
\usepackage{comment}
\usepackage{makecell}
\usepackage{braket}
\usepackage{placeins}
\usepackage{diagbox}
\usepackage{bbding}
\usepackage{ulem}
\usepackage{float}
\usepackage{afterpage}
\usepackage{dcolumn}
\usepackage{bm}
\usepackage[pdftex,colorlinks=true]{hyperref}
\usepackage[dvipsnames]{xcolor}
\definecolor{darkblue}{HTML}{1B436C}
\definecolor{darkred}{HTML}{94080A}
\hypersetup{
	colorlinks=true,
    urlcolor=darkred,
    citecolor=darkblue,
    linkcolor=darkblue
}
\definecolor{ForestGreen}{RGB}{34, 139, 34}

\newcommand{\affA}{State Key Laboratory of Artificial Microstructure and Mesoscopic
Physics, School of Physics, Peking University, Beijing 100871, China}
\newcommand{\affB}{Max Planck Institute for the Physics of Complex Systems, N\"othnitzer Str.~38, 01187 Dresden, Germany}
\newcommand{\affC}{Cavendish Laboratory, University of Cambridge, Cambridge CB3 0HE, United Kingdom}
\newcommand{\affE}{Technical University of Munich, TUM School of Natural Sciences,
Physics Department, 85748 Garching, Germany}
\newcommand{\affF}{Munich Center for Quantum Science and Technology (MCQST), Schellingstr. 4, 80799 M{\"u}nchen, Germany}

\begin{document}
\title{From stable periodic orbits to many-body chaos:\\ doubly tunable prethermalization via engineering of an emergent band structure}
\author{Jianan Wang}
\thanks{These authors contributed equally to this work.}
 \affiliation{\affA}
\author{Yang Hou}
\thanks{These authors contributed equally to this work.}
 \affiliation{\affA}
  \affiliation{\affB}
\author{Andrea Pizzi}
 \affiliation{\affC}
	\author{Johannes Knolle}
\affiliation{\affE}
\affiliation{\affF}
	\author{Roderich Moessner}
 \affiliation{\affB}
 	\author{Hongzheng Zhao}
	\email{hzhao@pku.edu.cn}
 \affiliation{\affA}
	\date{\today}

\begin{abstract}
We uncover a family of many-body periodic orbits in a periodically driven (Floquet) spin system away from the high-frequency limit. 
While linear
stability analysis predicts that perturbed many-body trajectories remain close to stable periodic orbits, thermodynamic principles dictate that Floquet heating will ultimately
set in. Our work aims to resolve the tension between these two expectations. In particular, we show that
perturbations away from the stable periodic orbits feature a description akin to a quasiparticle band structure. A long-lived prethermal regime appears when modes around the gapless point are {slowly}
populated. The dispersion determines the prethermal lifetime, and we show how  band engineering leads to a “doubly tunable” parametric dependence of the prethermal lifetime $R^{-W}$, {with $R$ the width in momentum space of the quasiparticle distribution} 
and $W$ the exponent of the dispersion around the gapless point. 
Our results not only establish a powerful route toward stabilizing non-equilibrium phases of matter in driven many-body systems {but also establish a conceptual bridge between periodic orbits in `low-dimensional' nonlinear systems and many-body chaos}.

\end{abstract}
\maketitle
\let\oldaddcontentsline\addcontentsline
\renewcommand{\addcontentsline}[3]{}

\textit{Introduction.---}Uncovering organizing principles governing the non-equilibrium behaviour of matter in interacting many-body systems is a central challenge of modern physics~\cite{d2016quantum}. 
For instance, time-dependent Hamiltonians can generate a wide variety of spatio-temporal orders, as in time crystals~\cite{khemani2016phase,else2016floquet,
yao2017discrete}. Yet, in the absence of energy conservation, driven many-body systems tend to absorb energy from the drive and heat up to featureless infinite-temperature states~\cite{Lazarides2014Equilibrium,Kim2014Testing,ponte2015periodically}.

A central question is therefore what mechanisms can stabilize non-equilibrium order and correlations against heating. By now, it is well established that the heating rate can be exponentially suppressed by a high-frequency drive, resulting in a long-lived prethermal regime described by an effective Hamiltonian before the eventual heat death~\cite{bukov2015universal,abanin2015exponentially,mori2016rigorous,zhao2019floquet,howell2019ssymptotic,rajak2019characterizations,else2020long,luitz2020prethermalization,rubio2020floquet,zhao2021random,ikeda2021fermi,pizzi2021classical,ye2021floquet,mori2022heating,jin2023fractionalized,fu2024engineeringhierarchicalsymmetries,zhang2026quantum}. 

Another promising way to suppress heating is to embed distinctive dynamical structures {involving only a few effective degrees of freedom} in many-body phase
space to constrain ergodic evolution~\cite{akila2017semiclassical,ho2019periodic,serbyn2021quantum,russomanno2023spatiotemporally,pizzi2025genuine,Theory2025Alessio,bhowmick2025granovskii,petrova2025finding,schmid2026genuine}. 
This idea has received sustained research attention, originating primarily from classical autonomous Hamiltonian systems.
For instance, stable periodic orbits (SPOs) in nonlinear systems can support nonergodic dynamics. A celebrated example is the Fermi-Pasta-Ulam-Tsingou chain
in which spatially localized breather states underpin long-lived metastable phenomena~\cite{fermi1955studies,flach2005q,flach2008discrete}. 
{In addition,
certain systems can be described by a low-dimensional effective mean-field description, such as in quantum quenches~\cite{LukinHuseSachdev_coarsening,Balducci_coarsening} or via the Gross–Pitaevskii equation, where SPOs have also been identified~\cite{Lerose2025Theory} and can support time crystalline behaviour~\cite{sacha2015modeling,pizzi2021classical}, which, although eventually unstable, can remain robust on experimentally accessible time scales. }

{While linear stability analysis predicts that perturbed many-body trajectories remain close to low-dimensional SPOs at early times, thermodynamic principles instead suggest that
Floquet heating will ultimately emerge at long times.
Our work aims to reconcile the tension between these two expectations and asks: What is the nature of this dynamical crossover? How can it be characterized, and how can it be tuned in a controlled manner?}

{We address these fundamental questions in a Floquet-driven many-body spin system and show that this dynamical crossover can be captured by an {effective harmonic description taking the form of a  
quasiparticle band structure, and for convenience
we will be using the language of quasiparticles in the following.} 
Remarkably, while complex nonlinear effects are fundamentally involved in the heating process, the quasiparticle band structure already within a {\it linearized} theory provides most valuable insights and even quantitative predictions.
In particular, a gapless dispersion gives rise to a long-lived prethermal regime, in which {the population of} low-energy quasiparticles slowly increases before the onset of a runaway heating process. As a key finding,  the less dispersive the occupied quasiparticles, the more delayed the heating.
} This results in a “doubly tunable” parametric dependence of the prethermal lifetime, $R^{-W}$, where $R$ characterizes the width of the initial quasiparticle distribution in momentum space and $W$ denotes the exponent of the dispersion near the gapless point. 

\begin{figure}[t]
    \centering
    \includegraphics[width=1\linewidth]{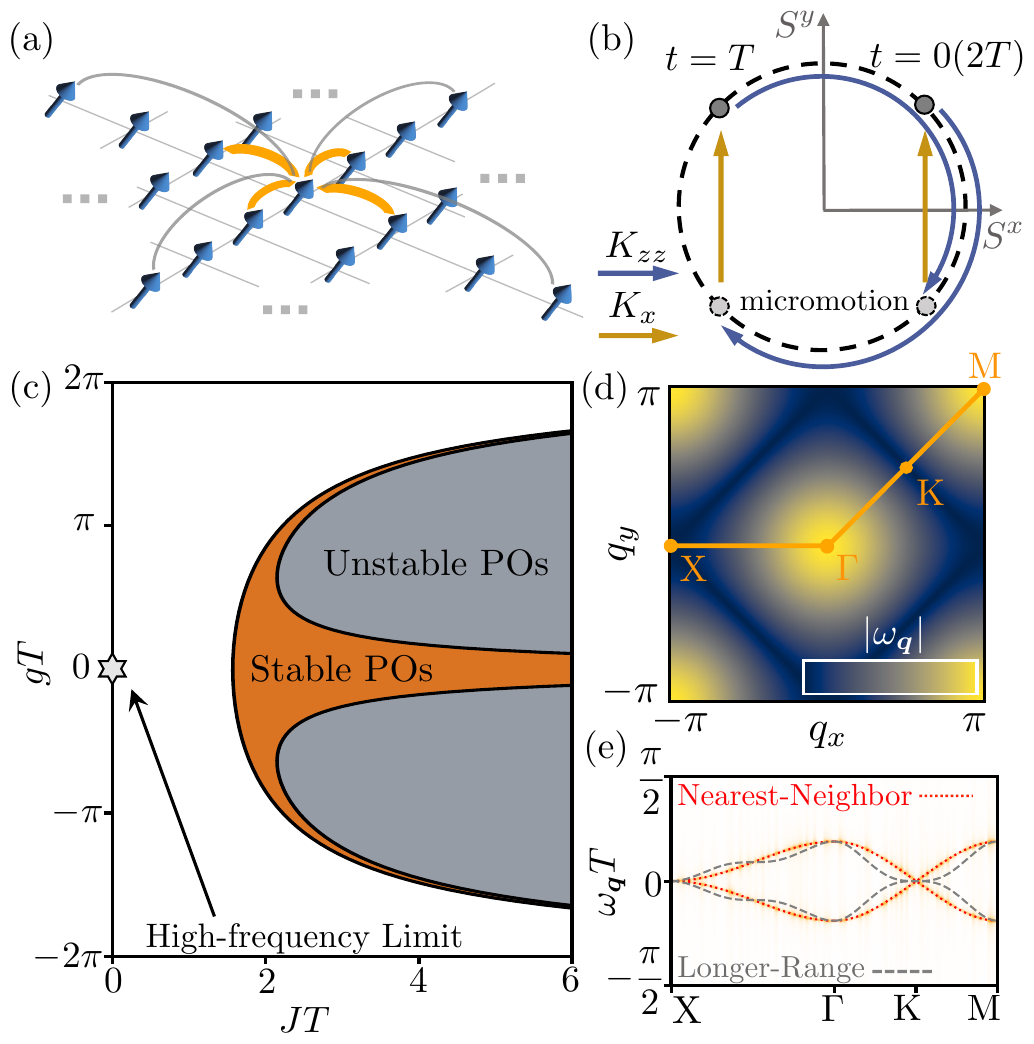}
    \caption{(a) Schematic of a 2D driven Ising model with nearest-neighbor (orange) and further-range (gray) interactions. (b) Spin dynamics projected on the $x-y$ plane follows a periodic orbit (PO), where the Ising interaction effectively echoes out the transverse field. 
    (c) Stability diagram of the POs in parameter space. SPOs exist in the orange region that is away from the high-frequency limit (star). (d) Dispersion  of the {quasiparticles}
    on top of SPOs. The spectrum is gapless, and closes along the dark blue diamond (including the \text{X} and \text{K} points). (e) Dispersion $\omega_{\boldsymbol{q}}$ along the path $\text{X}-\Gamma-\text{M}$. {Numerical results (orange)
    match analytical predictions (red dotted)}.
    Further-range interactions, as illustrated in (a), can be used to further flatten the band bottom around the $\text{K}$ point (gray). Numerical simulation in (e)
    uses $J = 1,\ g = 0.5,\ T = 2,\ L = 128,\ \sigma = 10^{-4}$.}
    \label{fig:PhaseDiagram}
\end{figure}

Concretely, for a Floquet-driven two-dimensional classical Ising system, we construct a family of SPOs beyond the high-frequency regime;
see Fig.~\ref{fig:PhaseDiagram}.
The quasiparticle description is obtained by linearizing the stroboscopic evolution around the SPOs, leading to a gapless dispersion, Fig.~\ref{fig:PhaseDiagram}(d) and (e). We further demonstrate prethermalization together with its parametrically tunable lifetime. Finally, we 
show that tailored further-range interactions can systematically flatten the band minima and further suppress heating, Fig.~\ref{fig:PhaseDiagram}(e).

Our results reveal the microscopic origin of the destruction of the SPOs and identify band-structure engineering as a powerful route toward stabilizing nonequilibrium phases of matter in driven systems.

\textit{Model and Periodic Orbits.---}We consider $L^2$ classical spins on a square lattice with periodic boundaries and a Hamiltonian $H(t)=H(t+T)$ with driving period $T$,
\begin{equation}\label{eq.Hamiltonian}
    H(t)=\begin{cases}
        H_1\equiv -\frac{J}{2}\sum_{\boldsymbol{r,a}}S^z_{\boldsymbol{r}}S^z_{\boldsymbol{r+a}}&\text{for }t\in[0,\frac{T}{2})\\[8pt]
        H_2\equiv -g\sum_{\boldsymbol{r}}S^x_{\boldsymbol{r}}&\text{for }t\in[\frac{T}{2},T)
    \end{cases},
\end{equation}
where $\boldsymbol{r}$ labels lattice sites and $\boldsymbol{a}=\pm\boldsymbol{\hat x},\pm\boldsymbol{\hat y}$ denotes lattice vectors. The normalized spin variables $\boldsymbol{S}_{\boldsymbol{r}}$ satisfy the Poisson bracket $\{S_{\boldsymbol{r}}^\mu,S_{{\boldsymbol{r}}'}^\nu\}=\delta_{{\boldsymbol{r}}{\boldsymbol{r}}'}\epsilon^{\mu\nu\rho}S_{\boldsymbol{r}}^{\rho}$, with the antisymmetric tensor $\epsilon^{\mu\nu\rho}$, and 
undergo Hamiltonian dynamics $\dot{S}_{\boldsymbol{r}}^\mu {=} \{S_{\boldsymbol{r}}^\mu, H(t)\}$. Here, $H_1$ comprises Ising interactions of strength $J$, inducing a nonlinear rotation $K_{zz}$ around the $z$ axis that depends on the average effective field of the neighboring spins; $H_2$ consists of a magnetic field and induces a uniform spin rotation $K_x$ about the $x$ axis with frequency $g$. {Both $K_{zz}$ and $K_{x}$ can be integrated exactly, and the evolved state over one driving period is obtained by applying discrete maps $\boldsymbol{S}_{\boldsymbol{r}}(T)=\left(K_x\circ K_{zz}\right)\boldsymbol{S}_{\boldsymbol{r}}(0)$.}

{Next, we adapt the state-selective spin echo technique recently introduced in Ref.~\cite{hou2025floquet} to construct {periodic orbits} (POs).
We consider the evolved state over two periods $\boldsymbol{S}_{\boldsymbol{r}}(2T)=\left(K_x\circ K_{zz}\circ K_x\circ K_{zz}\right)\boldsymbol{S}_{\boldsymbol{r}}(0)$. The key point is: if the rotation angles of both $K_{zz}$ are $\pi$, the rotations $K_x$ get echoed out, and the spins return to the initial states tracing out the closed orbit shown in Fig.~\ref{fig:PhaseDiagram}(b). Crucially, this echoing mechanism relies on the interaction and only works for initial states that, for every $\boldsymbol{r}$, satisfy the condition
\begin{equation}
\sum_{\boldsymbol{a}}S^z_{\boldsymbol{r+a}}(0)=\frac{2\pi}{JT},\quad \sum_{\boldsymbol{a}}S^y_{\boldsymbol{r+a}}(0)=\frac{2\pi}{JT}\tan\frac{gT}{4}.
\label{eq:constraint}
\end{equation}
This leads to a family of POs. For simplicity, in the following we focus on uniform states $\boldsymbol{S}_{\boldsymbol{r}} = \boldsymbol{S}_{\mathrm{PO}}$, with
\begin{equation}
\boldsymbol{S}_{\mathrm{PO}}{=}\frac{\pi}{2JT}\left[\sqrt{\left(\frac{2JT}{\pi}\right)^2{-}\sec^2{\frac{gT}{4}}},\ \tan{\frac{gT}{4}},\ 1\right]. 
\label{eq.PO_initial}
\end{equation}
Other POs can be constructed, e.g., by flipping the sign of ${S}^x_{\boldsymbol{r}}$; see one example in End Matter.}

\begin{figure*}[t]
    \centering
    \includegraphics[width=1\linewidth]{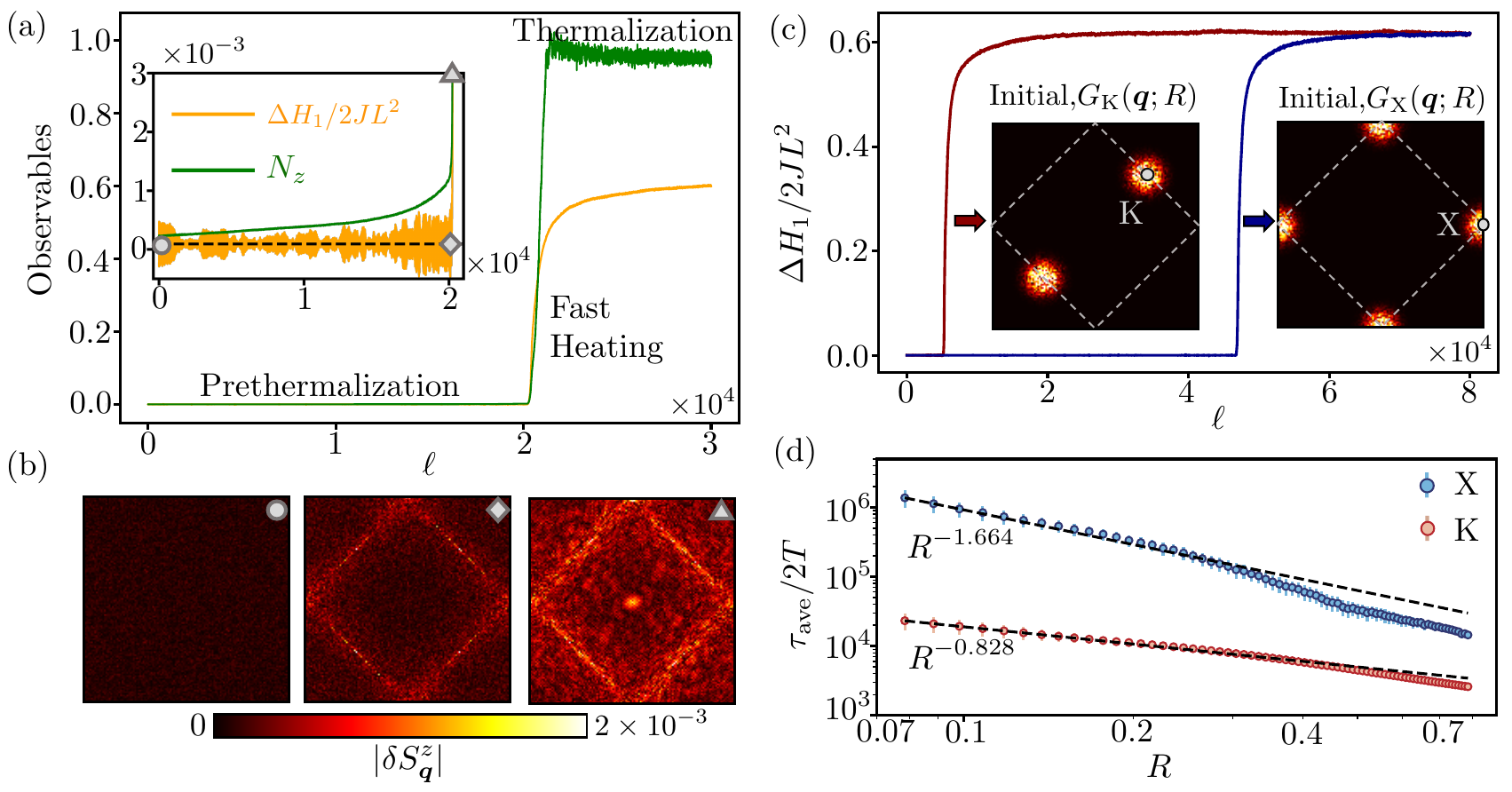}
    \caption{
    (a) Dynamics of total occupation number $N_z=\sum_{\boldsymbol{q}}|\delta S^z_{\boldsymbol{q}}|^2$ and change of Ising Hamiltonian $\Delta H_1/2JL^2$ from a perturbed SPO state. The system first enters a long-lived prethermal plateau due to the presence of the SPO until fast heating occurs at $\ell\approx2\times 10^4$. During the prethermal stage, the Ising energy {remains approximately constant} while the total occupation number undergoes a weak but continuous increase (inset). (b) Distribution of the quasiparticle occupation number in momentum space. {Initially, the distribution is random  (circle). As time evolves, new excitations accumulate around the gapless diamond (diamond). This is followed by a runaway heating process where other modes are also populated (triangle).}
    (c) Dynamics from localized Gaussian packets. The heating time of the initial configuration at the X point is much longer than that at the K point. (d) The prethermal lifetime of localized initial Gaussian packets is algebraically prolonged by reducing the width of the packets as $\tau\sim R^{-\alpha}$.
    Exponent $\alpha$ depends on the dispersion around the corresponding gapless point. 
    100 independent realizations are used for the average, and their standard deviation serves as the error bar. Numerical simulations are performed using $J=1$, $g=0.5$, $T=2$ and $L=128$, with $\sigma=0.0144$ for (a)(b) and $\sigma=0.3$ for (c)(d).
    }
    \label{fig:Dynamics}
\end{figure*}

The stability of POs can be assessed via linear stability analysis.
Concretely, we introduce a generic perturbation $\delta\boldsymbol{S_r}$ to the initial condition Eq.~\eqref{eq.PO_initial} and calculate its two-period stroboscopic evolution generated by Eq.~\eqref{eq.Hamiltonian}. This evolution can be linearized in $\delta\boldsymbol{S_r}$ and decoupled in  momentum space with a Fourier transform $\delta\boldsymbol{S_r}{=}\sum_{\boldsymbol{q}} \delta\boldsymbol{S_q}e^{i\boldsymbol{q\cdot r}}$, yielding $\delta\boldsymbol{S_q}(2T){=}\mathbf{M}_{\boldsymbol{q}}\cdot \delta\boldsymbol{S_q}(0)$, with a matrix $\mathbf{M_{\boldsymbol{q}}}$.

The eigenvalues of $\mathbf{M_{\boldsymbol{q}}}$ read $\exp\left[\left(\lambda_{\boldsymbol{q}}+\mathrm{i}\omega_{\boldsymbol{q}}\right)\cdot2T\right]$, with the Lyapunov exponent $\lambda_{\boldsymbol{q}}$ and the oscillatory frequency of the normal modes $\omega_{\boldsymbol{q}}$. The maximal Lyapunov exponent $\lambda_{\text{max}}{=}\max(\lambda_{\boldsymbol{q}})$ vanishes for $JTS^x_\mathrm{PO}\sin\frac{gT}{2}{\equiv}F_0{\leq} 1$, implying linear stability of the PO. In contrast, for $F_0{>}1$, $\lambda_{\text{max}}=\frac{1}{T}\mathrm{arccosh}F_0>0$ and perturbations grow exponentially in time, making the PO unstable and quickly leading to chaotic evolution; see details in Sec.~1 of the Supplementary Material (SM)~\cite{SM}.

The stability diagram of the POs in parameter space is shown in Fig.~\ref{fig:PhaseDiagram}(c), where $F_0=1$ sets the boundary separating stable (orange) and unstable (gray) regions. The numerically obtained phase boundary precisely matches the analytical prediction; see Fig.~S1 (SM)~\cite{SM}.
In some parameter regions (white) our method fails to construct POs, as it would require the spin in Eq.~\eqref{eq.PO_initial} to exceed unit length. We emphasize that inside such a region lies the regime $JT\to 0$: our POs are {\it not} tied to the conventional high-frequency regime.

\textit{Heating Dynamics.---} We turn to the full many-body dynamics obtained numerically. We prepare the system on an SPO, then perturb it {with $\delta S_{\boldsymbol{r}}^{y{/}z}$, where each site is independently sampled} from a Gaussian distribution with mean zero and standard deviation $\sigma$ (the $x$-component is then obtained according to the normalization constraint~\footnote{Note that some states from this initial sampling have spins exceeding unit length, and are therefore discarded.}). The resulting quasiparticle occupation number in momentum space is depicted in Fig.~\ref{fig:Dynamics}(b). 

Initially,
the perturbations around the SPOs
oscillate with frequency
\begin{equation}\label{eq.dispersion}
\omega_{\boldsymbol{q}}=\pm\arcsin(\gamma_{\boldsymbol{q}}F_0)/T,
\end{equation}
where $\gamma_{\boldsymbol{q}}{=}(\cos{q_x}{+}\cos{q_y})/2$. This dispersion is gapless for
$\gamma_{\boldsymbol{q}}=0$ on the dark blue diamond in Fig.~\ref{fig:PhaseDiagram}(c). Around the X-point, the dispersion is quadratic, whereas around the K-point we find a Dirac-like linear dispersion, see Fig.~\ref{fig:PhaseDiagram}(d). {Numerically, {we 
perform the space-time Fourier transform of the dynamics of $\delta{S}_{\boldsymbol{r}}^z$ at time $2\ell T$ over a long time window $0{<}\ell{\leq}300$.} 
The result (orange in Fig.~\ref{fig:PhaseDiagram}(d)) precisely follows our theoretical prediction (red dotted) and confirms the quasiparticle picture.} 
As we elaborate below, the dispersion plays a fundamental role in the heating of the system.

The system then stays in a long-lived prethermal regime, diagnosed by tracking the Ising Hamiltonian $H_1$. 
In Fig.~\ref{fig:Dynamics}(a), we present the stroboscopic dynamics of the change in the Ising energy for a single trajectory, $\Delta H_1{=}H_1(\ell \cdot 2T){-}H_{\text{PO}}$ (orange), where $H_{\text{PO}}$ denotes the Ising energy for the unperturbed PO. $\Delta H_1$ remains close to zero for a long time before a sudden increase happens around $\ell{\approx}2{\times}10^4$. This indicates that during this prethermal regime, $H_1$ {remains approximately constant}, as the driving protocol effectively echoes out the transverse field. 

This picture clarifies further in momentum space. We track the total quasiparticle occupation, $N_z=\sum_{\boldsymbol{q}}| \delta S^z_{\boldsymbol{q}}|^2$, and observe a weak (${\sim}10^{-3}$) but continuous increase in the prethermal regime; see the green curve in the inset of Fig.~\ref{fig:Dynamics}(a). This process corresponds to an interesting microscopic dynamics in momentum space, Fig.~\ref{fig:Dynamics}(b), where gapless modes with $\gamma_{\boldsymbol{q}}{=}0$ (diamond in momentum space) are slowly populated. {It arises as the system evolves towards higher entropy {while keeping $H_1$ approximately unchanged}: in momentum space, $H_1$ can be rewritten as $  H_1{=}{-}2J L^2\left((S^z_{\text{PO}})^2{+}2S^z_{\text{PO}}\delta S^z_{\boldsymbol{q=0}}{+}\sum_{\boldsymbol{q}}\gamma_{\boldsymbol{q}}|\delta S^z_{\boldsymbol{q}}|^2\right)$, and populating 
near-gapless modes contributes negligibly to the Ising energy~\footnote{Note that when ${\boldsymbol{q}}=0$, $\gamma_{\boldsymbol{q}}$ is nonzero. Hence, excitations around ${\boldsymbol{q}}=0$ will change the value of $H_1$ and hence $\delta S^z_{\boldsymbol{q=0}}$ can only oscillate around 0 slightly.}.}

\textit{Doubly Tunable Prethermal Lifetime.---} {After the long-lived prethermal regime, the system heats up to infinite temperature, Fig.~\ref{fig:Dynamics}(a). Such heating is triggered by complex nonlinear effects: 
When multiple modes are populated, different values of  $\gamma_{\mathbf{q}}$, which quantifies how fast at stroboscopic times the modes oscillate around SPOs, cause relative phases to accumulate; in real space, this results in a growing mismatch between the spin configurations of each mode, which in turn facilitates the generation of further nonlinearities. This directly leads to the exponential growth of the total occupation number $N_z \sim \exp(\kappa t)$ with growth rate $\kappa$, until a local heating channel triggers uncontrollable heating; see details in the End Matter.
Importantly, when the system is initialized near a gapless point of the spectrum such that $\gamma_{\mathbf{q}}$ is narrowly distributed, the accumulation of relative phases is suppressed, and so is the rate $\kappa$.}

To demonstrate this, we focus on two special gapless points, X with $\boldsymbol{q}_\text{X}=(\pi,0)$ and K with $\boldsymbol{q}_\text{K}=(\pi/2,\pi/2)$. We prepare our initial states as follows. We first perturb SPOs with Gaussian random variables $ \delta S^{y{/}z}_{\boldsymbol{q}}$, then apply a Gaussian packet envelope $G_{\text{K/X}}(\boldsymbol{q};R)$ with width $R$ to localize the initial modes at one of the gapless points~\footnote{As a convention, we fix the initial value of $\sum_{\boldsymbol{q}}| \delta S_{\boldsymbol{q}}^z|^2$ by normalizing the envelope when varying $R$.}; see insets of Fig.~\ref{fig:Dynamics}(c). Details of the envelope function are provided in  Sec.~2 in the SM~\cite{SM}. As $R$ is reduced to zero, fewer modes are occupied, and heating is suppressed. Interestingly, as shown in Fig.~\ref{fig:Dynamics}(c), for the same value of $R$, the state centered around X heats up one order of magnitude more slowly than that centered around K. 
Quantitatively, as shown in Fig.~\ref{fig:Dynamics}(d), for small $R$, the prethermal lifetime is algebraically prolonged as $R^{-\alpha}$. Numerical fitting suggests that $\alpha_\text{X}\approx 2$ and $\alpha_{\text{K}}\approx 1$, implying a strong dependence on the spectrum structure around the corresponding gapless point.

\begin{figure}[t]
    \centering
    \includegraphics[width=1\linewidth]{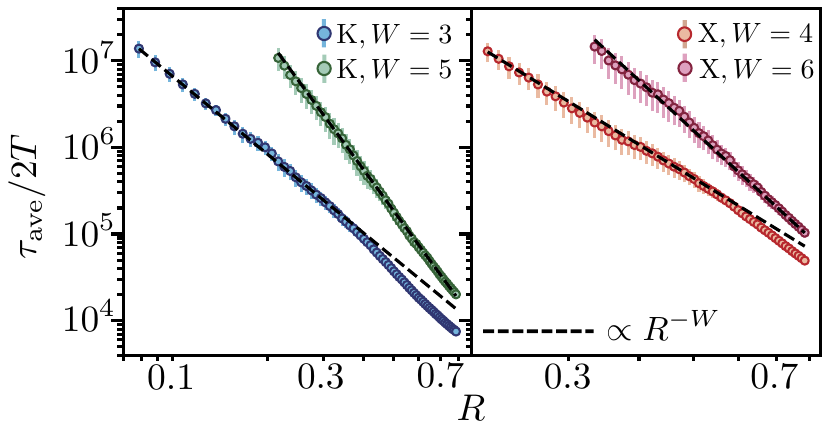}
    \caption{Prethermal lifetime $\tau_\mathrm{ave}$ versus initial distribution width $R$ for engineered band structures with $W=3,4,5,6$. The numerical results match well with the theoretical scaling $R^{-W}$,  plotted as a black dashed line as a guide to the eye. 100 independent realizations are used for the average, and their standard deviation serves as the error bar. We use $\sigma=0.176$ with
    $J=1$, $g=1$, $T=2$, and $L=128$ for the  simulations. }
    \label{fig:Lifetime}
\end{figure}
{We explain this scaling through a perturbative analysis. Near the gapless point where $\gamma_{\boldsymbol q}$ is small, we expand the stroboscopic equation of motion to the leading order in $\gamma_{\mathbf{q}}$ and to the quadratic order in the perturbation amplitude, $\mathcal{O}(|\delta S|^2)$, to capture the dominant non-linear effects. {We focus on the excitation rate of the quasiparticle occupation number $N_z$. By treating $\delta S^{y,z}_{\boldsymbol{r}}$ as Gaussian 
random variables, we  estimate how nonlinear effects accumulate as time evolves, and find
(see Sec.~4 of the SM~\cite{SM})}
\begin{equation}\label{eq:GrowthEquation}
\frac{\mathrm d}{\mathrm d t}\left\langle N_z\right\rangle\propto R^W\langle N_z\rangle,
\end{equation}
where $W$ characterizes the scaling exponent of the dispersion around a gapless point $\boldsymbol{q_0}$, 
\begin{equation}
\gamma_{\boldsymbol{q}}\sim|\boldsymbol{q-q_0}|^W.
\label{eq.dispersionscaling}
\end{equation} Here $\langle\cdot\rangle$ denotes the time average over a time window that is much longer than the typical oscillatory period of the linear modes but smaller than the prethermal lifetime. 
Eq.~\eqref{eq:GrowthEquation} then implies a lifetime $\tau\sim R^{-W}$.}

{This argument predicts $\alpha_\text{K}=1,\ \alpha_\text{X}=2$. However, in practice, quasiparticles can scatter away from the gapless point such that Eq.~\eqref{eq.dispersionscaling} no longer holds perfectly, and the initial Gaussian wave packet distribution in momentum space becomes distorted as time evolves. Both effects can cause the fitted exponents to be smaller than the ideal values, as shown in Fig.~\ref{fig:Dynamics} (d).}

The dispersion around the gapless point can be tuned to more general algebraic forms, thereby enabling extra control of heating. For instance, as shown in Fig.~\ref{fig:PhaseDiagram}(d), we can incorporate further-ranged Ising couplings between sites separated by odd lattice spacing
\begin{equation}
\label{eq.long-range}
    H_1^{[M]}{=}{-}\frac{J}{2}\sum_{\boldsymbol{r,a}}\sum_{m=1}^{M}c_mS^z_{\boldsymbol{r}}S^z_{\boldsymbol{r}{+}(2m{-}1)\boldsymbol{a}},
\end{equation}
with $c_m$  dimensionless coefficients. Requiring $\sum_mc_m{=}1$ ensures that this construction preserves the specific form of POs given by the initial condition, Eq.~\eqref{eq.PO_initial}. $\gamma_{\boldsymbol{q}}$ can now be engineered through an appropriate choice of the coefficients $c_m$. Concretely, we impose $\nabla_{\boldsymbol{q}}^k\gamma_{\boldsymbol{q}}^{[M]}=0$ for $k=1,2,\cdots, W-1$ at the gapless points, yielding a tunable exponent $W\leq2M$. In particular, $W=2M-1$ can be  realized around the K point by setting
\begin{equation}
c_m^\text{{K}}
=
\frac{1}
{2^{2M-2}}\binom{2M-1}{M-m},
\end{equation}
where $\binom{n}{k}=\frac{n!}{k!(n-k)!}$ represents the binomial coefficients. Similarly, $W=2M$ is  obtained at the X point with 
\begin{equation}
c_m^\text{{X}}
=
\frac{(-1)^{m-1}M}
{2^{4M-3}(2m-1)}\binom{2M}{M}\binom{2M-1}{M-m}.
\end{equation}
We further numerically verify the scaling for $W=3,4,5,6$, as shown in Fig.~\ref{fig:Lifetime}. The results match the theoretical prediction well when the momentum distribution is narrow with a small value of $R$.

\textit{Discussion.---}
We have investigated the long-time stability of SPOs in a driven many-body spin system. Fluctuations around the SPOs can be described as quasiparticle excitations with a gapless dispersion. Before the onset of runaway heating, the system first enters a long-lived prethermal regime where the gapless modes are slowly populated. 

{We show how to systematically suppress heating by {flattening the dispersion around the gapless
point
via tailored further-range interactions.} In general, these interactions tend to accelerate heating in the high-frequency regime as they enlarge the local energy scale and open more heating channels
~\cite{kuwahara2016floquet,machado2020long,moon2025experimental}. However, we explicitly show how this can be avoided and discover the doubly tunable prethermalization lifetime scaling $R^{-W}$. More broadly, this scaling reveals a dynamical manifestation of higher-order van Hove singularities~\cite{classen2025high} in non-equilibrium driven systems. Importantly, although the driving now involves interactions beyond {nearest-neighbour interactions,} 
Eq.~\eqref{eq.long-range} remains local and can be efficiently implemented using state-of-the-art quantum simulators}~\cite{bernien2017probing,hollerith2022realizing,Google2025ErrorCorrection,weckesser2025realization}. {Also, the initial state under consideration can be prepared via single-site rotations}~\cite{weitenberg2011single,liu2026prethermalization}.

Our method thus presents an unprecedented and experimentally feasible scheme to stabilize non-equilibrium phases of matter in driven systems far away from the high-frequency regime. 
This idea should generally apply to SPOs in other driven systems. 
Note, the band structure also strongly depends on the lattice geometry and the system dimension. Using these ingredients to further tailor the band structures, e.g., to open a gap or generate topological bands, is worth pursuing.

For numerical efficiency, here we have focused on classical systems,
but we anticipate that the key heating suppression mechanism can be generalized to
quantum many-body systems.
It remains an interesting open question whether, and how, quantum fluctuations affect the dispersion relation and if they can lead to fundamentally different heating phenomena. Also, the existence of SPOs can constrain the entanglement growth, thereby enabling large-scale numerical simulations of quantum many-body dynamics using tensor-network techniques. Our current Floquet protocol preserves discrete time-translational symmetry, and the existence and long-time stability of SPOs in quasi-periodically~\cite{zhao2019floquet,else2020long,PhysRevLett.120.070602,long2022many,Critically2025Pilatowsky,zhou2026emergent} or even randomly driven systems~\cite{zhao2021random,wen2021periodically} remains largely unexplored. Finally, questions arise regarding the effect of dissipative or contractive dynamics~\cite{KPZ2025Daviet,Ethan2026,eckstein2026dynamical,yoshida2026theory} to protect the SPOs.

\textit{Acknowledgment.---}
This work is supported 
by Quantum Science and Technology-National Science and Technology Major Project
(No. 2024ZD0301800) and
by the National Natural Science
Foundation of China (Grant No. 12474214), and by the High-performance Computing Platform of Peking University.
AP acknowledges support by Trinity College Cambridge.
This research was supported in part by grant NSF PHY-2309135 to the Kavli Institute for Theoretical Physics (KITP), {as well as by the Deutsche Forschungsgemeinschaft under grants FOR 5522 (project-id 499180199) and the cluster of excellence ctd.qmat (EXC 2147, project-id 390858490).}

\bibliography{ref}
\clearpage
\part*{\scalebox{0.6}{End Matter}}
\addcontentsline{toc}{part}{End Matter} 
\textit{Microscopic Heating Mechanism.---}We now provide a detailed analysis of the microscopic heating mechanism. In the main text, we observe that during the heating process, momentum modes around the gapless points are slowly excited. Although this {does not change the Ising energy $H_1$ perceptibly}, it can open a detrimental local heating channel that triggers fast heating.

\begin{figure}[h]
    \centering
    \includegraphics[width=1\linewidth]{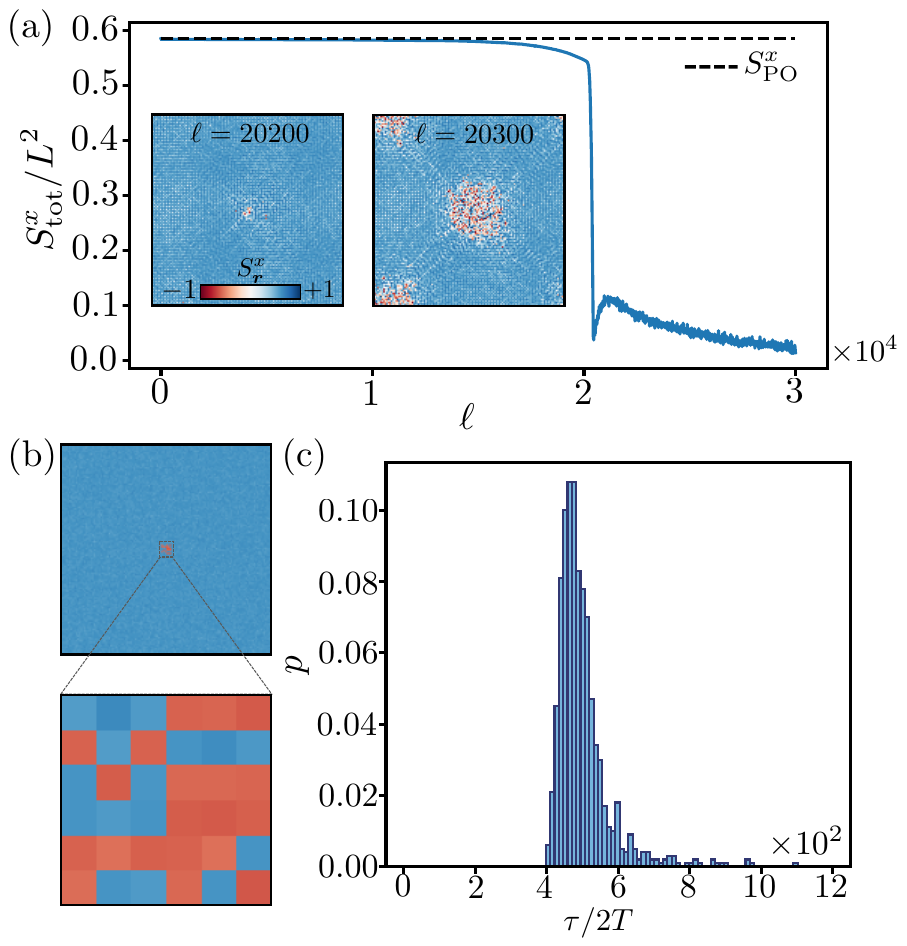}
    \caption{(a) Dynamics of the total magnetization in the $x$ direction, which continuously decreases during prethermalization. The insets are snapshots of configurations of $S^x_{\boldsymbol{r}}$, where a locally unstable region appears with flipped $S^x_{\boldsymbol{r}}$ (red). This region quickly expands and triggers fast heating. (b) The initial state where the signs of $S^x_{\boldsymbol{r}}$ within a $6\times6$ square region are randomly chosen. (c) Distribution of the prethermal lifetime from the initial states in (b). The local flip significantly shortens the prethermal lifetime, from $\mathcal{O}(10^4)$ to $\mathcal{O}(10^2)$. 1000 independent realizations are used to sample the distribution. Numerical simulations are performed using $J=1,g=0.5,T=2,L=128$ and $\sigma=0.0144$.}
    \label{fig:xdroplet}
\end{figure}

To show this, we begin by monitoring the dynamics of the $x$-spin component. Fig.~\ref{fig:xdroplet}(a) shows the dynamics of the change in total magnetization along the $x$ direction. While the total occupation number grows continuously, as observed in the main text in Fig.~\ref{fig:Dynamics}(a), the $x$-magnetization also decreases during the prethermal stage. We further examine the real-space spin configurations of the $x$-component at time points right before the fast heating, shown in the inset of Fig.~\ref{fig:xdroplet}(a). It is clear that the locally unstable region with flipped $S^x_{\boldsymbol{r}}$ spontaneously emerges from the background and subsequently grows. Note, such local heating channels have also been recently reported in Refs.~\cite{hou2025floquet,guo2025emergent}.
In fact, these flips are a consequence of continuous excitation of gapless modes during prethermalization. As detailed in Sec.~2 in the SM~\cite{SM}, we show that the decay of the $x$-component of the total magnetization is proportional to the total occupation number $N_z$ at early times. The increase in $N_z$ leads to a continuous decay of $S^x_{\mathrm{tot}}$. Microscopically, this is manifested as local flips of $S^x_{\boldsymbol{r}}$. 

This spatial non-uniformity, even in a local region, can trigger fast heating, and we show it from the perspective of POs. {We begin from the uniform state $\boldsymbol{S}_{\boldsymbol{r}} = \boldsymbol{S}_{\mathrm{PO}}$ and introduce local flips of $S^x_{\boldsymbol{r}}\to-S^x_{\boldsymbol{r}}$. Note, this operation does not break the POs since Eq.~\eqref{eq:constraint} can still be perfectly satisfied after this flipping.} However, we find that these POs can be unstable against perturbations once the $x$-components substantially deviate from a ferromagnetic state. As a solvable example, the maximal Lyapunov exponent of the antiferromagnetic alignment $S^x_{\boldsymbol{r}=(i,j)}=(-1)^{i+j}S^x_\text{PO}$ is $\lambda_{\mathrm{max}}=\text{arcsinh}F_0/T>0$, corresponding to unstable POs whenever $F_0$ is non-vanishing. Therefore, local flips of $S^x_{\boldsymbol{r}}$ can induce chaotic dynamics that destroys the prethermal order. We further check the argument above by evolving the system from the perturbed states where the sign of $S^x_{\boldsymbol{r}}$ within a $6\times6$ subregion is randomly chosen, as shown in Fig.~\ref{fig:xdroplet}(b). Fig.~\ref{fig:xdroplet}(c) demonstrates the distribution of the prethermal lifetime for 1000 realizations with the above initial states. Most trajectories heat up at a time of $\mathcal{O}(10^2)$, which is two orders of magnitude shorter than that of SPOs, which is around $\mathcal{O}(10^4)$.

Therefore, we conclude that, starting from a uniformly perturbed SPO, the non-linear effects lead to a continuous growth in the total occupation number $\sum_{\boldsymbol{q}}|\delta S^z_{\boldsymbol{q}}|^2$. It further reduces the total magnetization in the $x$-direction and opens a local heating channel that triggers rapid heating.

\clearpage

\let\addcontentsline\oldaddcontentsline
	\cleardoublepage
	\onecolumngrid
 \begin{center}
\textbf{\large{\textit{Supplementary Material} \\ \smallskip
	From stable periodic orbits to many-body chaos:\\
doubly tunable prethermalization via engineering of an emergent band structure}}
    \\[1em]
Jianan Wang, $^{1,*}$ Yang Hou, $^{1,2,*}$ Andrea Pizzi, $^{3}$ Johannes Knolle, $^{4,5}$ Roderich Moessner, $^{2}$ Hongzheng Zhao, $^{1,\dagger}$
\\[.1cm]
{\small
$^{1}$\textit{\affA} \\
$^{2}$\textit{\affB} \\
$^{3}$\textit{\affC} \\
$^{4}$\textit{\affE} \\
$^{5}$\textit{\affF}\\
}
(Dated: \today)\\[1cm]
	\end{center}
	\renewcommand{\thefigure}{S\arabic{figure}}
        \setcounter{figure}{0}
    \renewcommand{\thesection}{SM\;\arabic{section}}
	\setcounter{section}{0}
	\renewcommand{\theequation}{S.\arabic{equation}}
        \setcounter{equation}{0}
    \renewcommand{\thesubsection}{\arabic{subsection}}
	\setcounter{section}{0}
    \tableofcontents
    \setcounter{page}{1}

\section{Periodic Orbits and Linear Stability Analysis}
Here we present the details of constructing periodic orbits (POs) and linear stability analysis when they are weakly perturbed.
\label{SM:Lyapunov}
\subsection{Construction of POs}
The time evolution of the system is governed by the Hamiltonian Eq.~\eqref{eq.Hamiltonian}. 
The equations of motion (EOMs) can be generated via $\dot{S}_{\boldsymbol{r}}^\mu {=} \{S_{\boldsymbol{r}}^\mu, H(t)\}$, which can be analytically integrated in our case.  During the first half-period, the time evolution follows the non-linear rotation $K_{zz}$ about the $z$-axis:
\begin{equation}\label{eq.mapzz}
    K_{zz}(\boldsymbol{S}_{\boldsymbol{r}})=\begin{pmatrix}
        S_{\boldsymbol{r}}^x\cos(\chi_{\boldsymbol{r}}T/2)+S_{\boldsymbol{r}}^y\sin(\chi_{\boldsymbol{r}}T/2)
        \\
        S_{\boldsymbol{r}}^y\cos(\chi_{\boldsymbol{r}}T/2)-S_{\boldsymbol{r}}^x\sin(\chi_{\boldsymbol{r}}T/2)\\
        S_{\boldsymbol{r}}^z
    \end{pmatrix},
\end{equation}
with the rotating frequency determined by the average $z$-magnetization of nearest-neighbor spins $\chi_{\boldsymbol{r}}^z=J\sum_{\boldsymbol{a}}S^z_{\boldsymbol{r+a}}$, where $\boldsymbol{a}=\pm\boldsymbol{\hat x},\pm\boldsymbol{\hat y}$ runs through the lattice vectors. The dynamics in the second half-period follows the rotation $K_x$ about the $x$-axis:
\begin{equation}
    K_{x}(\boldsymbol{S}_{\boldsymbol{r}})=\begin{pmatrix}
        S_{\boldsymbol{r}}^x\\
        S_{\boldsymbol{r}}^y\cos(gT/2)-S_{\boldsymbol{r}}^z\sin(gT/2)
        \\
        S_{\boldsymbol{r}}^z\cos(gT/2)+S_{\boldsymbol{r}}^y\sin(gT/2)
    \end{pmatrix}.
\end{equation}
The evolved state over one driving period is obtained as $\boldsymbol{S}(T)=\left(K_x\circ K_{zz}\right)\boldsymbol{S}(0)$.

We consider the evolution over two periods 
\begin{equation}
\label{eq.twocycle}
    \boldsymbol{S}(2T)=\left(K_x\circ K_{zz}\circ K_x\circ K_{zz}\right)\boldsymbol{S}(0).
\end{equation} The state-selective spin echo requires that the rotation angles by both $K_{zz}$ are $\pi$, and then the rotations by $K_x$ are effectively echoed out~\cite{hou2025floquet}.  As a result, the spins return to the initial states tracing out the closed orbit. Concretely, this requires the state to satisfy the condition
\begin{equation}
    \frac{JT}{2}\sum_{\boldsymbol{a}}S^z_{\boldsymbol{r+a}}(0)=\frac{JT}{2}\sum_{\boldsymbol{a}}S^z_{\boldsymbol{r+a}}(T)=\pi,
\end{equation}
which turns into constraints on the initial conditions:
\begin{equation}
    \sum_{\boldsymbol{a}}S^z_{\boldsymbol{r+a}}(0)=\frac{2\pi}{JT},\quad \sum_{\boldsymbol{a}}S^y_{\boldsymbol{r+a}}(0)=\frac{2\pi}{JT}\tan\frac{gT}{4}.
\end{equation}
Note that there are various choices of initial states that satisfy the constraints for the POs. For simplicity, we consider uniform states as shown in Eq.~\eqref{eq.PO_initial}.

\subsection{Linear Stability Analysis}
We now perform linear stability analysis on these POs. We slightly perturb the POs as $\boldsymbol{S_r}=\boldsymbol{S}_\mathrm{PO}+\delta\boldsymbol{S_r}$ with $|\delta\boldsymbol{S_r}|\ll1$. Substituting the perturbed state into the EOM Eq.~\eqref{eq.twocycle} and keeping the leading-order terms in $\mathcal{O}(|\delta\boldsymbol{S_r}|)$, we obtain the linearized dynamics. {With periodic boundary conditions}, this can be decoupled by Fourier transform $\delta\boldsymbol{S_r}{=}\sum_{\boldsymbol{q}} \delta\boldsymbol{S_q}e^{i\boldsymbol{q\cdot r}}$. For each $\boldsymbol{q}-$mode, the decoupled linear map reads $\delta\boldsymbol{S_q}(2T){=}\mathbf{M}_{\boldsymbol{q}}\cdot \delta\boldsymbol{S_q}(0)$, with the Jacobian matrix 
\begin{equation}
    \boldsymbol{M_q}=\begin{pmatrix}
        1&-\gamma_{\boldsymbol{q}}Y\sin\theta&\gamma_{\boldsymbol{q}}Y(1-\cos\theta)+\gamma^2_{\boldsymbol{q}}YX\sin\theta\\
        0&1-\gamma_{\boldsymbol{q}}X\sin\theta\cos\theta&-\gamma_{\boldsymbol{q}}X(1+\cos^2\theta)+\gamma^2_{\boldsymbol{q}}X^2\cos\theta\sin\theta\\
        0&\gamma_{\boldsymbol{q}}X\sin^2\theta&1+\gamma_{\boldsymbol{q}}X\sin\theta\cos\theta-\gamma^2_{\boldsymbol{q}}X^2\sin^2\theta
    \end{pmatrix},
    \label{finalmatrix}
\end{equation}
where $X=\sqrt{(2JT)^2-\pi^2/\cos^2\frac{gT}{4}},\ Y=\pi\tan\frac{gT}{4},\ \theta=\frac{gT}{2}$, and $\gamma_{\boldsymbol{q}}=(\cos q_x+\cos q_y)/2$. $\gamma_{\boldsymbol{q}}$ encodes the lattice in Fourier space
\begin{equation}\label{eq.dispersionfactor}
    \frac{JT}{2}\sum_{\boldsymbol{a}}\delta S^z_{\boldsymbol{r+a}}=2JT\sum_{\boldsymbol{q}}\gamma_{\boldsymbol{q}}\delta S^z_{\boldsymbol{q}}\exp(\mathrm{i}\boldsymbol{q\cdot r}).
\end{equation}
It controls both the linear stability and the nonlinear effects (see Sec.~\ref{SM:Lifetime}) and plays an important role in engineering the band structure (see Sec.~\ref{SM:Engineering}).  

The eigenvalues $\Lambda_{\boldsymbol{q}}$ of the Jacobian matrix $\boldsymbol{M_q}$ read
\begin{equation}
    \Lambda_{\boldsymbol{q}}=1,-\left(\frac{\gamma_{\boldsymbol{q}}}{2}X\sin\theta+\sqrt{\frac{\gamma^2_{\boldsymbol{q}}}{4}X^2\sin^2\theta-1}\right)^2,-\left(\frac{\gamma_{\boldsymbol{q}}}{2}X\sin\theta-\sqrt{\frac{\gamma^2_{\boldsymbol{q}}}{4}X^2\sin^2\theta-1}\right)^2.
\end{equation}
Writing  $\Lambda_{\boldsymbol{q}}=\exp((\lambda_{\boldsymbol{q}}+\mathrm{i}\omega_{\boldsymbol{q}})\cdot2T)$ yields the Lyapunov exponents
\begin{equation}
    \lambda_{\boldsymbol{q}}=\begin{cases}
        0,\pm\text{arccosh }|\gamma_{\boldsymbol{q}}F_0|/T,\quad &|\gamma_{\boldsymbol{q}}F_0|>1\\[8pt]
        0~(\text{three-fold}), & |\gamma_{\boldsymbol{ q}}F_0|\leq1
    \end{cases},
\end{equation}
and the oscillatory frequency of the normal modes
\begin{equation}
    \omega_{\boldsymbol{q}}=\begin{cases}
        0,\pm{\pi}/{(2T)},\quad  &|\gamma_{\boldsymbol{q}}F_0|>1\\[8pt]
     0,\pm\arcsin(\gamma_{\boldsymbol{q}}F_0)/T, \quad &|\gamma_{\boldsymbol{q}}F_0|\leq1
    \end{cases},
    \label{eq:dispersion}
\end{equation}
where $F_0=JT\sin\frac{gT}{2}S^x_{\mathrm{PO}}$. The linear stability of the POs is then determined by the maximal Lyapunov exponent $\lambda_{\text{max}}{=}\max(\lambda_{\boldsymbol{q}})$. For $F_0{\leq} 1$, $\lambda_{\text{max}}$ vanishes, leading to SPOs where the perturbation does not grow in time. Instead, these perturbations feature oscillatory motion with a dispersion  $\omega_{\boldsymbol{q}}=\pm\arcsin(\gamma_{\boldsymbol{q}}F_0)/T$. In contrast, for $F_0{>}1$, $\lambda_{\text{max}}=\frac{1}{T}\mathrm{arccosh}F_0>0$, leading to unstable periodic orbits where the perturbation grows exponentially in time. This leads to chaotic evolution and the many-body system rapidly becomes featureless. 

\subsection{Numerical Verification}
\begin{figure}[h]
    \centering
    \includegraphics[width=0.65\linewidth]{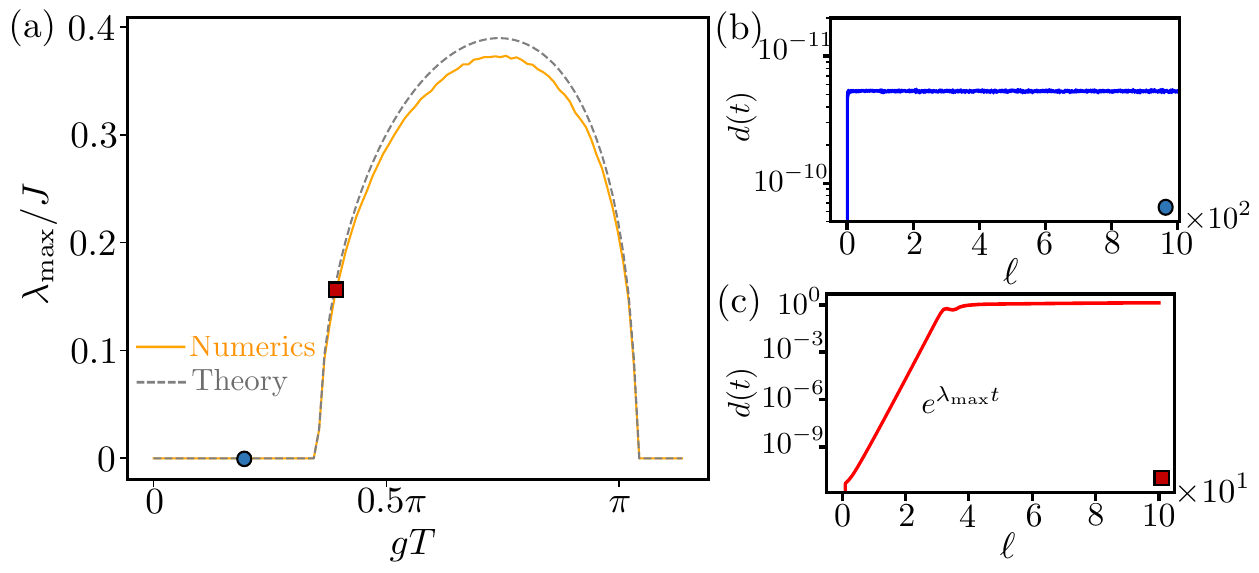}
    \caption{(a) Maximal Lyapunov exponent $\lambda_\text{max}$ versus $gT$. The growth rate of the amplitude of the perturbation can be used to estimate the Lyapunov exponent, as shown in (b)(c). The numerical results (orange) match well with the analytical calculation (gray). (b)(c) Dynamics of $d(t)$ to show the difference between $\lambda_{\text{max}}=0$ and $\lambda_{\text{max}}\neq0$. We set $g=0.08\pi$ in (b) (stable PO), $g=0.16\pi$ in (c) (unstable PO). We also fix $T = 2.5,\ \sigma=10^{-12},\ L=128,\ J=1$ for numerical simulations.}
    \label{fig:Lyapunov}
\end{figure}
Now we present the numerical verification of the linear stability analysis.
The short-time dynamics are well captured by the linearized evolution. We numerically compute the amplitude of the perturbation $d(t){=}\sqrt{\sum_{\boldsymbol{r}}|\delta\boldsymbol{S_r}(t)|^2/L^2}$ . For unstable POs, this grows exponentially at short times, $d(t){\sim}d(0)e^{\lambda_\mathrm{max}t}$, where the growth rate corresponds to the Lyapunov exponent (see Fig.~\ref{fig:Lyapunov}(b)(c)). In Fig.~\ref{fig:Lyapunov}(a), we fix $JT=2.5$ and scan over a wide range of $g$ and plot the numerical Lyapunov exponents (orange). The numerics aligns well with the theoretical prediction of maximal Lyapunov exponents, covering both stable and unstable regimes.  Note, since the initial perturbation occupies all momentum modes, smaller values of $\lambda_{\boldsymbol{q}}$ also contribute to the dynamics. Consequently, the numerically fitted results can be slightly smaller than the analytical prediction. 

\begin{figure}[h]
    \centering
    \includegraphics[width=0.6\linewidth]{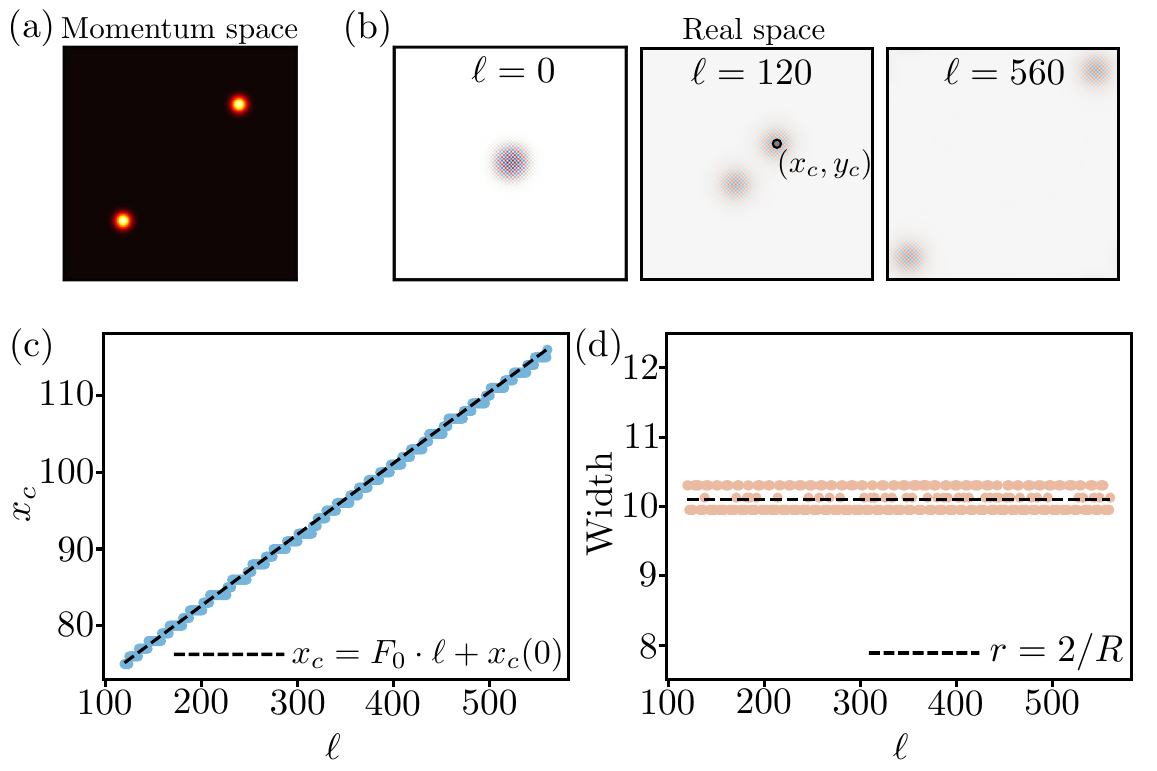}
    \caption{(a) Initial state distribution of $| \delta S^z_{\boldsymbol{q}}|$ in the momentum space as a Gaussian packet. (b) Real-space distribution $\delta S^z_{\boldsymbol{r}}$ at different times, where the ballistic propagation of the wave packet is clearly observed. In the middle panel, $(x_c,y_c)$ parameterizes its position and $y_c=x_c$ because the wave packet moves along the diagonal direction. (c) Position of the packet center versus the driving cycles $\ell$, with the speed obeying the analytical prediction (black). (d) The width of the distribution remains approximately unchanged as time evolves. Numerical simulation is performed using $J=1,g=1.332,T=2,L=128,R=\pi/16,A=10^{-10}$.}
    \label{fig:Ballistic}
\end{figure}

Perturbations around SPOs oscillate in time, following the dispersion relation as shown in the main text, Fig.~\ref{fig:PhaseDiagram} (e).
In addition, one can also predict interesting wave packet dynamics using this dispersion relation. In static systems, a Gaussian wave packet centered at the momentum $\boldsymbol{q}$ propagates ballistically with the group velocity $\nabla_{\boldsymbol{q}}\omega_{\boldsymbol{q}}$, while its profile spreading is governed by the second derivative $|\nabla^2_{\boldsymbol{q}}\omega_{\boldsymbol{q}}|$. The emergent dispersion in our systems also leads to a similar dynamical phenomenon. For instance, we prepare a Gaussian packet at $\boldsymbol{q}_\mathrm K$ as $\delta S^z_{\boldsymbol{q}}=A(e^{-\mathrm (\boldsymbol{q-q}_\text{K})^2/R^2}+e^{-\mathrm (\boldsymbol{q+q}_\text{K})^2/R^2})$, as shown in Fig.~\ref{fig:Ballistic}(a), and monitor its early-time dynamics before the onset of heating. The dispersion predicts the group velocity 
\begin{equation}
    \boldsymbol{v_g}(\boldsymbol{q_\text{K}})=\frac{\partial\omega_{\boldsymbol{q}}}{\partial \boldsymbol{q}}|_{\boldsymbol{q=q_{\text{K}}}}=\pm\frac{F_0}{2T}\left(\hat x+\hat y\right),
\end{equation}
and the second-order derivative vanishes. This thus leads to ballistic propagation without spreading, as shown in Fig.~\ref{fig:Ballistic}(b), where the wave packet propagates along the diagonal direction as expected. The position of the packet versus time is plotted in Fig.~\ref{fig:Ballistic}(c), where a clear linear dependence is observed. The velocity matches precisely the theoretical group velocity, i.e., $x_c(2\ell T)-x_c(0)=v_g\cdot 2\ell T=\ell\cdot F_0$, which is drawn as a black dashed line as a guide to the eye. In addition, during the evolution, the width of the wave packet in real space remains almost unchanged, as shown in Fig.~\ref{fig:Ballistic}(d). 

\section{Preparation of Localized Initial States}
\label{sec:localized}
We provide details of the initial state preparation. In the main text, {we first sample $\delta S^z_{\boldsymbol{r}}$ on each site independently from a Gaussian distribution with standard deviation $\sigma$.}  Then we use the envelope function $G_{\text{K/X}}(\boldsymbol{q}; R)$ to localize the distribution in momentum space. The explicit form of the function reads
\begin{equation}
    G_{\text{K}}(\boldsymbol{q};R)=g(\boldsymbol{q}-(\frac{\pi}{2},\frac{\pi}{2});R)+g(\boldsymbol{q}-(-\frac{\pi}{2},-\frac{\pi}{2});R)\ ;
\end{equation} 
\begin{equation}
    G_{\text{X}}(\boldsymbol{q};R)=g(\boldsymbol{q}-(\pi,0);R)+g(\boldsymbol{q}-(0,\pi);R),
\end{equation} 
where $g(\boldsymbol{q-q_c};R)$ is a Gaussian function centered at $\boldsymbol{q_c}$ with width $R$ 
\begin{equation}
    g(\boldsymbol{q-q_c};R)=\frac{R_0}{R}\exp(-\frac{(\boldsymbol{q-q_c})^2}{R^2}),
\end{equation}
where $R_0$ is a normalization constant. We set $R_0=5\pi/32$ for numerical simulations.  This envelope function ensures that the classical spins are real vectors, and the total quasiparticle occupation number is fixed for different $R$, i.e., $N_z(0)\approx\sum_{\boldsymbol{q}}G_{\text{K/X}}^2(\boldsymbol{q};R)\sigma^2/L^2\approx R^2_0\sigma^2/4\pi$.

\section{Dynamics of $x/y$-components}
\label{SM:XY-Dynamics}
In the main text, we only focus on the $z-$component of $\boldsymbol{\delta S_q}$, as it is closely related to the Ising energy $H_1$. In this section, we discuss the dynamics of other components. 
\begin{figure}[h]
    \centering
    \includegraphics[width=1\linewidth]{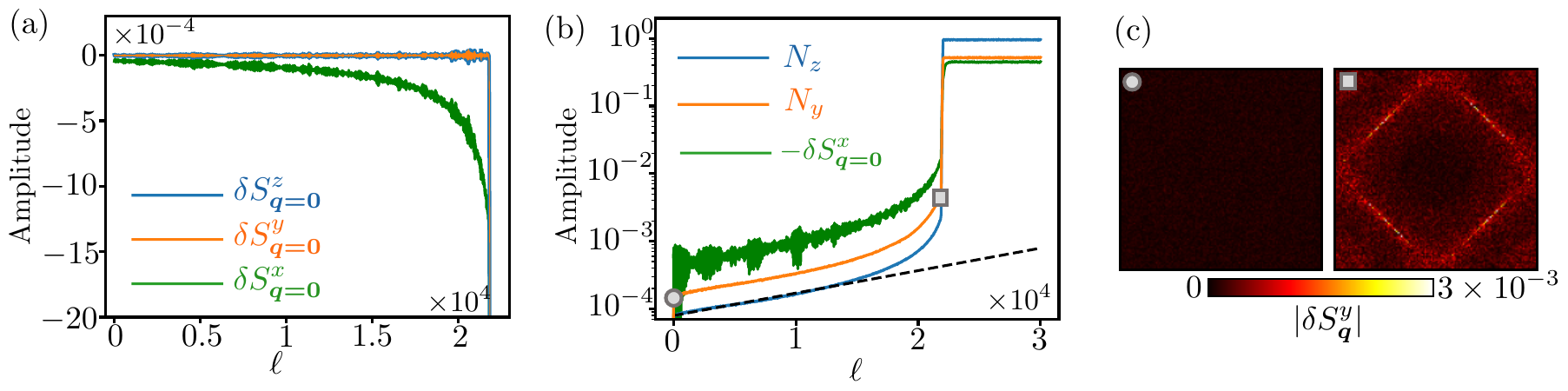}
    \caption{(a) Dynamics of $\delta S^{x,y,z}_{\boldsymbol{q=0}}$. At early times, the total magnetization in the $z$ and $y$ directions remains unchanged, while the $x$-magnetization decreases continuously. (b) Dynamics of $N_{y,z}$ and $-\delta S^x_{\boldsymbol{q=0}}$. The black dashed line fits the early-stage exponential growth of $N_z$. (c) Snapshots of configurations in $\boldsymbol{q}-$space of the $y-$component. From a uniformly distributed state (circle), the newly excited momentum modes spontaneously accumulate around the gapless diamond (square). Numerical simulations are performed using $J=1$, $g=1$, $T=2$, $L=128$ and $\sigma=0.008$.
    }
    \label{fig:XYZDynamics}
\end{figure}

In Fig.~\ref{fig:XYZDynamics}(a), we track the dynamics of the total magnetization along all three directions $\sum_{\boldsymbol{r}}S^\mu_{\boldsymbol{r}}/L^2=\delta S^\mu_{\boldsymbol{q=0}}$, $\mu=x,y,z$. During the prethermal stage, both $y$- and $z$-directions slightly oscillate around 0, while the $x$-magnetization decreases continuously. The total occupation number associated with the $y$-component, $N_y=\sum_{\boldsymbol{q}}|\delta S^y_{\boldsymbol{q}}|^2$, also increases slowly (note that $N_x$ is not independent of $N_y$ and $N_z$ due to the normalization condition) after a short transient, as shown in Fig.~\ref{fig:XYZDynamics}(b).  Indeed, the growth rates for $N_y$ and $N_z$ are approximately the same. We use a log scale in Fig.~\ref{fig:XYZDynamics}(b) and a linear fit (black dashed line) suggests that this growth is exponential as a function of time at early times. 

Further, as shown in Fig.~\ref{fig:XYZDynamics}(b), $-\delta S^x_{\boldsymbol{q=0}}$ (green) shows a similar dynamical behavior to $N_y$ and $N_z$. To understand this, we use the constraint of the unit length, $(S^x_{\text{PO}}+\delta S^x_{\boldsymbol{r}})^2+(S^y_{\text{PO}}+\delta S^y_{\boldsymbol{r}})^2+(S^z_{\text{PO}}+\delta S^z_{\boldsymbol{r}})^2=1$ and show 
\begin{equation}
    \delta S^x_{\boldsymbol{q=0}}=\sum_{\boldsymbol{r}}\delta S^x_{\boldsymbol{r}}/L^2=-\frac{S^y_{\text{PO}}\delta S^y_{\boldsymbol{q=0}}+S^z_{\text{PO}}\delta S^z_{\boldsymbol{q=0}}}{S^x_{\text{PO}}}-\sum_{\boldsymbol{k}}\frac{|S^y_{\text{PO}}\delta S^y_{\boldsymbol{k}}+S^z_{\text{PO}}\delta S^z_{\boldsymbol{k}}|^2/(S^x_{\text{PO}})^2+|\delta S^y_{\boldsymbol{k}}|^2+|\delta S^z_{\boldsymbol{k}}|^2}{2S^x_{\text{PO}}}+\mathcal{O}(|\delta \boldsymbol S|^3),
    \label{eq:X_0}
\end{equation}
Given the numerical observation that $y$- and $z$- components exhibit similar dynamics at early times, as well as the fact that $S^{y,z}_{\boldsymbol{q=0}}\approx 0$, we obtain
\begin{equation}
     \label{eq:Sx-Nz}S^x_{\boldsymbol{q=0}}\propto N_z,
\end{equation}
where we neglect the higher-order contributions $\mathcal{O}(|\delta \boldsymbol S|^3)$. Eq.~\eqref{eq:Sx-Nz} will be important in the following section when we derive the doubly tunable prethermal lifetime.

\section{Doubly Tunable Prethermal Lifetime}
\label{SM:Lifetime}
We now provide a detailed analysis for the doubly tunable prethermal lifetime $\tau\sim R^{-W}$ by studying the dominant non-linear effects. 

\subsection{Derivation of the Lifetime Scaling $\tau\sim R^{-W}$}

The heating process is governed by nonlinearities which induce scattering between quasiparticles and create new excitations. We focus on the growth rate of quasiparticles, i.e., the rate of the change of $N_z=\sum_{\boldsymbol{q}}| \delta S^z_{\boldsymbol{q}}|^2$. The prethermal lifetime can then be estimated by the inverse of the growth rate.

According to Sec.~\ref{SM:XY-Dynamics}, at the early stage of the evolution, we have $N_z\approx-\nu\delta S^x_{\boldsymbol{q=0}}$, where $\nu$ is a positive constant. Thus, the growth rate can be obtained as 
\begin{equation}\label{eq:ratenoave}
    \frac{\mathrm{d}}{\mathrm{d}t}N_z\approx -\nu\frac{\mathrm{d}}{\mathrm{d}t}\delta S^x_{\boldsymbol{q=0}}=-\frac{\nu}{L^2}\sum_{\boldsymbol{r}}\frac{\mathrm{d}}{\mathrm{d}t}\delta S^x_{\boldsymbol{r}}=-\frac{\nu}{L^2}\frac{1}{2T}\sum_{\boldsymbol{r}}{S^x_{\boldsymbol{r}}((\ell+1)\cdot 2T)-S^x_{\boldsymbol{r}}(\ell\cdot 2T)}.
\end{equation}
Due to the non-linearity of the system, it is challenging to make analytical progress for generic initial conditions. However, for initial states that are prepared around the gapless point, we can perturbatively expand the right-hand side by keeping only terms that are linear in $\gamma_{\boldsymbol{q}}$, which is now a small parameter. In addition, we also only keep the contributions that are quadratic in the perturbation amplitude $|\delta \boldsymbol{S}|$. We then arrive at
\begin{equation}\label{eq:nonlinear}
    \begin{split}
        \delta S^x_{\boldsymbol{r}}((\ell+1)\cdot 2T)-\delta S^x_{\boldsymbol{r}}(\ell\cdot 2T)&\approx \text{Linear Terms}\\&+2JT\left[\Delta _{\boldsymbol{r}}^{z}(\ell\cdot 2T)(\cos^2{\frac{gT}{2}}+1)+\Delta _{\boldsymbol{r}}^{y}(\ell\cdot 2T)\sin{\frac{gT}{2}}\cos{\frac{gT}{2}} \right]\delta S^y_{\boldsymbol{r}}(\ell\cdot 2T)\\&+2JT\left[-\Delta _{\boldsymbol{r}}^{z}(\ell\cdot 2T)\sin{\frac{gT}{2}}\cos{\frac{gT}{2}}+\Delta _{\boldsymbol{r}}^{y}(\ell\cdot 2T)\sin^2{\frac{gT}{2}}\right]\delta S^z_{\boldsymbol{r}}(\ell\cdot 2T),
    \end{split}
\end{equation}
where 
\begin{equation}
    \Delta^{z/y}_{\boldsymbol r}
    \equiv \sum_{\boldsymbol a}
    \delta S^{z/y}_{\boldsymbol r+\boldsymbol a}=
    \sum_{\boldsymbol q}
    \gamma_{\boldsymbol q}
    \delta S^{z/y}_{\boldsymbol q}
    e^{i\boldsymbol q\cdot\boldsymbol r},
\end{equation}
are terms linear in both $\gamma_{\boldsymbol{q}}$ and $|\delta\boldsymbol{S}|$. This is a complicated expression that involves non-linear effects. To proceed, 
we assume that, in accordance with the numerical observation (see Sec.~\ref{SM.EGP}), the quasiparticle distribution in momentum space remains close to the initial Gaussian distribution. Then Eq.~\eqref{eq:nonlinear} can be simplified as an integral over Gaussian random variables that depend both on $N_z$ and the width of the initial sampling $R$. 
We further recast these variables to separate their explicit dependence on $N_z$ and $R$, which, as elaborated below, allows us to directly extract the growth rate.

The initial perturbations $ \delta S^{y,z}_{\boldsymbol{r}}$ are sampled from a Gaussian distribution with zero mean and standard deviation $\sigma$. Note, Gaussian variables remain Gaussian under any linear transformation, and hence the Fourier components $ \delta S^{y,z}_{\boldsymbol{q}}$ are now complex Gaussian variables with zero mean and standard deviation $\sigma/L$, subject to a constraint $\delta S^{y,z}_{\boldsymbol{-q}}=(\delta S^{y,z}_{\boldsymbol{q}})^*$ which ensures that $\delta S^{y,z}_{\boldsymbol r}$ is real. We then apply a Gaussian wave packet envelope $G_{\text{X/K}}(\boldsymbol{q};R)$; see Sec.~\ref{sec:localized}. Since multiplying a Gaussian variable by a deterministic factor simply rescales its standard deviation, this procedure is equivalent to directly sampling $\delta S^{y,z}_{\boldsymbol{q}}$ from a complex Gaussian distribution with zero mean and a $\boldsymbol{q}$-dependent standard deviation $G_{\text{X/K}}(\boldsymbol{q};R)\sigma/L$, subject to a constraint $\delta S^{y,z}_{\boldsymbol{-q}}=(\delta S^{y,z}_{\boldsymbol{q}})^*$. Using the variance-addition property of Gaussian random numbers, $\delta S^{y,z}_{\boldsymbol{r}}$ follows a Gaussian distribution with zero mean and a variance of 
\begin{equation}
    \sum_{\boldsymbol{q}}(G_{\text{X/K}}(\boldsymbol{q};R)\sigma/L)^2= N_z.
\end{equation}
$\Delta^{y,z}_{\boldsymbol{r}}$ follows a Gaussian distribution with zero mean and a variance of \begin{equation}\label{eq:Mismatch}
    \sum_{\boldsymbol{q}}\left(\gamma_{\boldsymbol{q}}^{[M]}G_{\text{X/K}}(\boldsymbol{q};R)\sigma/L\right)^2\approx
    \frac{\sigma^2R^2_0}{L^2}\int\frac{\left|\boldsymbol{q}-\boldsymbol{q}_{\mathrm{X/K}}\right|^{2W}}{R^2}\exp\left(-\frac{2\left(\boldsymbol{q}-\boldsymbol{q}_{\mathrm{X/K}}\right)^2}{R^2}\right)\mathrm{d}^2\boldsymbol{q}\propto N_zR^{2W},
\end{equation}
where $W$ characterizes the scaling exponent of the dispersion around a gapless point, $\gamma_{\boldsymbol{q}}\sim\mathcal{O}(|\boldsymbol{q}-\boldsymbol{q}_\mathrm{X/K}|^W)$. We can therefore parametrize these random variables as
\begin{equation}
    \delta S^{y,z}_{\boldsymbol{r}}=\sqrt{N_z}\epsilon^{y,z}_{\boldsymbol{r}},\quad \Delta^{y,z}_{\boldsymbol{r}}=R^W\sqrt{N_z}\eta^{y,z}_{\boldsymbol{r}},
    \label{eq:Substitution}
\end{equation}
where $\epsilon^{y,z}_{\boldsymbol r}$ and $\eta^{y,z}_{\boldsymbol r}$ are rescaled random variables whose variances tend to be largely independent of $N_z$ and $R$. Substituting Eq.~\eqref{eq:nonlinear} and Eq.~\eqref{eq:Substitution} into Eq.~\eqref{eq:ratenoave}, we obtain
\begin{equation}
    \frac{\mathrm{d}}{\mathrm{d}t}N_z=\text{Linear Terms }+N_zR^W\cdot\frac{\nu J}{L^2}\sum_{\boldsymbol{r}}\left[(\epsilon^z_{\boldsymbol{r}}\eta^z_{\boldsymbol{r}}-\epsilon^y_{\boldsymbol{r}}\eta^y_{\boldsymbol{r}})\cos\frac{gT}{2}\sin\frac{gT}{2}-2\epsilon^z_{\boldsymbol{r}}\eta^y_{\boldsymbol{r}}\cos^2\frac{gT}{2}\right].
\end{equation}
The linear terms correspond to oscillations around the SPOs. These oscillations can be averaged out by applying a time average $\langle\cdot\rangle$ over a time window which needs to be much larger than the typical oscillatory period of the linear modes but smaller than the prethermal lifetime. The resulting time-averaged excitation rate equation reads
\begin{equation}\label{eq:HeatingRate}
    \frac{\mathrm{d}}{\mathrm{d}t}
    \left\langle N_z\right\rangle\approx N_zR^W\times\frac{\nu J}{L^2}\sum_{\boldsymbol{r}}\left\langle\left[(\epsilon^z_{\boldsymbol{r}}\eta^z_{\boldsymbol{r}}-\epsilon^y_{\boldsymbol{r}}\eta^y_{\boldsymbol{r}})\cos\frac{gT}{2}\sin\frac{gT}{2}-2\epsilon^z_{\boldsymbol{r}}\eta^y_{\boldsymbol{r}}\cos^2\frac{gT}{2}\right]\right\rangle=\Phi \left\langle N_z\right\rangle R^W,
\end{equation}
where
\begin{equation}
   \label{eq.phi}
   \Phi=\frac{\nu J}{L^2}\sum_{\boldsymbol{r}}\left\langle\left[(\epsilon^z_{\boldsymbol{r}}\eta^z_{\boldsymbol{r}}-\epsilon^y_{\boldsymbol{r}}\eta^y_{\boldsymbol{r}})\cos\frac{gT}{2}\sin\frac{gT}{2}-2\epsilon^z_{\boldsymbol{r}}\eta^y_{\boldsymbol{r}}\cos^2\frac{gT}{2}\right]\right\rangle.
\end{equation}
Crucially, we numerically verify that $\Phi$ is positive and largely independent of $R$ and $N_z$ in Sec.~\ref{SM.phi}. The prethermal lifetime is obtained by the inverse of the excitation rate, leading to 
\begin{equation}
    \tau\propto R^{-W}.
    \label{eq:tau_R_N}
\end{equation}
{On the square lattice and when only nearest-neighbor interactions are present, $W$ can only take on values 1 or 2.} Other values of $W$ can be realized through band engineering by introducing {further-range interactions}; see details in Sec.~\ref{SM:Engineering}. {In addition, it remains an interesting question to investigate the possibility of tuning $\Phi$
 to further control heating. We leave this question for future work.}

\subsection{Quasiparticle Distribution Function}\label{SM.EGP}
In the previous section, we assumed that the quasiparticle distribution
in momentum space remains close to its initial distribution. We now verify this assumption numerically.

\begin{figure}[h]
    \centering
    \includegraphics[width=1\linewidth]{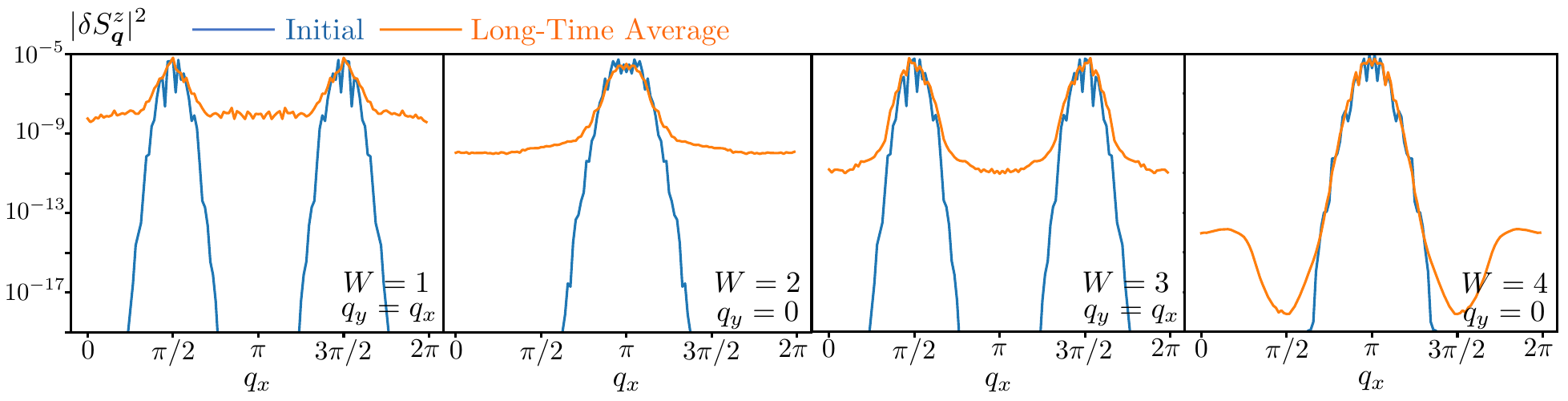}
    \caption{Comparison between the initial quasiparticle distribution(blue) and the time-evolved distribution(orange). We choose the momentum $q_y=q_x$ for odd $W$ and $q_y=0$ for even $W$. As $W$ increases, extra excitation can be suppressed. Numerical simulations are performed using $J=1,\ g=1,\ T=2,\ L=128,\ R=3\pi/32$ and $\sigma=0.176$.}
    \label{fig:LocalDynamics}
\end{figure}

We compare the quasiparticle momentum-space distribution in the initial states and the time-evolved states. 
Fig.~\ref{fig:LocalDynamics} plots the distribution for models with $W=1,2,3,4$. Note that, for time-evolved states, we perform a time average over a long time window that spans three-quarters of the prethermal lifetime to reduce temporal fluctuations. 
We choose $q_y=q_x$ for odd $W$ and $q_y=0$ for even $W$. The time-evolved distribution (orange) indeed stays close to the initial distribution (blue). However, weak excitations that are not occupied in the initial states can appear. However, as $W$ increases, the occupation of these excitations can be suppressed from $10^{-8}$ to $10^{-14}$ as $W$ increases from 1 to 4, such that this assumption becomes more valid. 
{We can also compare the numerical results in Fig.~\ref{fig:Dynamics}(d) and Fig.~\ref{fig:Lifetime}. The fitted scaling exponent $\alpha_\mathrm {X/K}$ indeed matches well with the theoretical prediction when $W$ is large.}

\subsection{Numerical Verification of $\Phi$}\label{SM.phi}
Another assumption that we made in the previous derivation is that the coefficient $\Phi$ in Eq.~\eqref{eq:HeatingRate} is positive and independent of $N_z$ and $R$. Here, we verify this numerically.

\begin{figure}[h]
    \centering
    \includegraphics[width=0.55\linewidth]{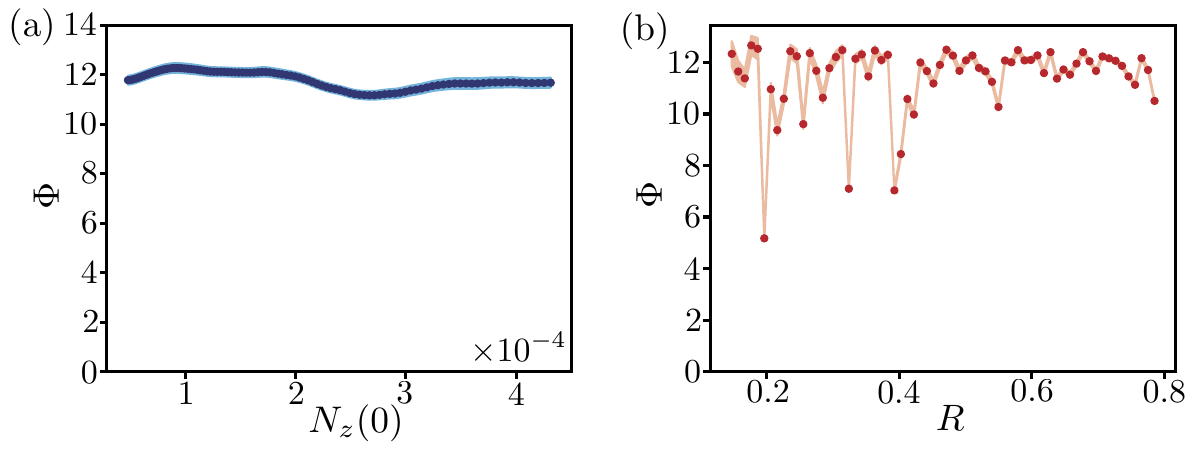}
    \caption{(a) $\Phi$ versus $N_z(0)$ when fixing $R=3\pi/32$. (b) $\Phi$ versus $N_z(0)$ when fixing $\sigma=0.05$ or $N_z(0)=4.8\times 10^{-5}$. $\Phi$ is positive and it is largely independent of $R$ and $N_z(0)$.  Numerical simulations are performed using $J = 1,\ g = 1,\ T = 2,\ L = 128,\ W=3$.}
    \label{fig:Phi}
\end{figure}

We consider initial states with different values of $N_z(0)$ and $R$ and perform the numerical simulation to extract $\Phi$, which is depicted in Fig.~\ref{fig:Phi}. Clearly, $\Phi$ is positive and barely changes with $N_z(0)$ and $R$.

\section{Engineering the Structure of the Dispersion Spectrum}
\label{SM:Engineering}
Heating around the SPOs in our system features a doubly tunable prethermal lifetime, which depends on the dispersion scaling exponent near the gapless point. Here we show how to systematically increase this scaling exponent by incorporating further-range Ising couplings.

We include further-range Ising coupling between sites separated by an odd lattice spacing in addition to the nearest-neighbor interaction 
\begin{equation}
    H_1^{[M]}=-\frac{J}{2}\sum_{\boldsymbol{r,a}}\sum^M_{m=1}c_mS^z_{\boldsymbol{r}}S^z_{\boldsymbol{r}+(2m-1)\boldsymbol{a}},
\end{equation}
where $c_m$ are dimensionless coefficients. The evolution generated by the above Hamiltonian still follows Eq.~\eqref{eq.mapzz} with a different  frequency $\chi_{\boldsymbol{r}}=J\sum_{\boldsymbol{a}}\sum_{m=1}^M c_mS^z_{\boldsymbol{r}+(2m-1)\boldsymbol{a}}$. By requiring $\sum_{m=1}^Mc_m=1$, we ensure that the further-range interaction still generates a $\pi$-rotation around the $z$-axis from the uniform state in Eq.~\eqref{eq.PO_initial}. Therefore, the system respects the specific form of the POs introduced in the main text. Furthermore, the phase boundary separating the stable and unstable regions is also unchanged. The dispersion relation of the perturbation on top of SPOs reads 
\begin{equation}
\omega_{\boldsymbol{q}}=\pm\arcsin(\gamma_{\boldsymbol{q}}^{[M]}F_0)/T,
\end{equation}
with a different $\gamma_{\boldsymbol{q}}^{[M]}$ from Eq.~\eqref{eq.dispersion},
\begin{equation}
    \gamma_{\boldsymbol{q}}^{[M]}=\frac{1}{2}\sum_{m=1}^Mc_m\{\cos[(2m-1)q_x]+\cos[(2m-1)q_y]\}.
\end{equation}
This is obtained from the Fourier transform of the perturbed rotation angle about the $z$-axis
\begin{equation}
    \frac{JT}{2}\sum^M_{m=1}\sum_{\boldsymbol{a}}c_m \delta S^z_{\boldsymbol{r}+(2m-1)\boldsymbol{a}}=2JT\sum_{\boldsymbol{q}}\exp(\mathrm{i}\boldsymbol{q\cdot r})\gamma_{\boldsymbol{q}}^{[M]}\delta S_{\boldsymbol{q}}^z,
\end{equation}
cf. Eq.~\eqref{eq.dispersionfactor}. The dispersion  $\gamma_{\boldsymbol{q}}^{[M]}$ can now be engineered through an appropriate choice of the coefficients $c_m$. Notice that both X and K points are gapless in this new dispersion; we thus determine the coefficients $c_m$ by requiring 
\begin{equation}
\nabla_{\boldsymbol{q}}^k\gamma_{\boldsymbol{q}}^{[M]}=0,\quad k=1,2,\cdots, W-1,
\end{equation} 
at these gapless points. Note that the above equations, together with the normalization condition on the coefficients, constitute a linear problem for solving $c_m$. For an exponent $W<2M-1$, this system admits infinitely many solutions. In particular, for $W=2M-1$ around the K point and $W=2M$ around the \(X\) point, the linear system has full rank, and the solution is unique. We now provide the solution for these two specific cases.

\textbf{{(i)}. $W=2M$ around the X point}

Around the X point, $\gamma_{\boldsymbol{q}}^{[M]}$ is cosinusoidal, for which the odd-order derivatives vanish automatically. In order to ensure that even-order derivatives up to $2M$ also vanish, we require
\begin{equation}
\sum_{m=1}^{M}c_m x_m^n=\delta_{n0},\quad x_m=(2m-1)^2,
\quad n=0,1,\dots,M-1.
\end{equation}
This is equivalent to a Vandermonde-type linear system
\begin{equation}
\begin{pmatrix}
1 & 1 & \cdots & 1\\
x_1 & x_2 & \cdots & x_M\\
x_1^2 & x_2^2 & \cdots & x_M^2\\
\vdots & \vdots & \ddots & \vdots\\
x_1^{M-1} & x_2^{M-1} & \cdots & x_M^{M-1}
\end{pmatrix}
\begin{pmatrix}
c_1\\
c_2\\
c_3\\
\vdots\\
c_M
\end{pmatrix}
=
\begin{pmatrix}
1\\
0\\
0\\
\vdots\\
0
\end{pmatrix}.
\label{eq:cmX}
\end{equation}
Since the nodes $x_m=(2m-1)^2$ are pairwise distinct, the Vandermonde matrix is invertible, and the solution is thus unique. The solution can be obtained by using Lagrange interpolation as
\begin{equation}
    c_m^\text{{X}}
=
\frac{(-1)^{m-1}M}
{2^{4M-3}(2m-1)}\binom{2M}{M}\binom{2M-1}{M-m}.
\end{equation}
Particularly, to obtain the results for $W=4$ in Fig.~\ref{fig:Lifetime} in the main text, we use
\begin{equation}
c_1=\frac{9}{8},\quad c_2=-\frac{1}{8},
\end{equation} 
for numerical simulations; for $W=6$, we use
\begin{equation}
c_1=\frac{75}{64},\quad c_2=-\frac{25}{128},\quad c_3=\frac{3}{128}.
\end{equation} 

\textbf{{(ii)}. $W=2M-1$ around the K point}

Around the K point, $\gamma_{\boldsymbol{q}}^{[M]}$ is sinusoidal, such that even-order derivatives are zero. By requiring odd-order derivatives up to $2M-1$ to vanish, we obtain
\begin{equation}
\sum_{m=1}^M(-1)^mc_m(2m-1)^{2n-1}=0,\quad n=1,\dots,M-1.
\end{equation}
By defining $y_m=(2m-1)^2$ and $\eta_m=(-1)^m(2m-1)c_m$, the above equation becomes a Vandermonde null-vector problem
\begin{equation}
\begin{pmatrix}
1 & 1 & \cdots & 1\\
y_1 & y_2 & \cdots & y_M\\
y_1^2 & y_2^2 & \cdots & y_M^2\\
\vdots & \vdots & \ddots & \vdots\\
y_1^{M-2} & y_2^{M-2} & \cdots & y_M^{M-2}
\end{pmatrix}
\begin{pmatrix}
\eta_1\\
\eta_2\\
\eta_3\\
\vdots\\
\eta_M
\end{pmatrix}
=\boldsymbol{0}.
\end{equation}

Since the nodes $y_m=(2m-1)^2$ are pairwise distinct, this null space dimension is one. A standard null vector is
\begin{equation}
\eta_m=
\frac{C}{\displaystyle\prod_{\substack{j=1\\ j\neq m}}^{M}(y_m-y_j)},
\end{equation}
where $C$ is a constant. Choosing $C$ such that $\sum_m c_m=1$, we obtain the solution to this linear system
\begin{equation}
c_m^\text{K}
=
\frac{(2M-1)!}
{4^{M-1}(M-m)!(M+m-1)!}=\frac{1}
{2^{2M-2}}\binom{2M-1}{M-m}
.
\end{equation}
Particularly, for $W=3$ in Fig.~\ref{fig:Lifetime} in the main text, we use
\begin{equation}
c_1=\frac{3}{4},\quad c_2=\frac{1}{4},
\end{equation} and for $W=5$ we use
\begin{equation}
c_1=\frac{5}{8},\quad c_2=\frac{5}{16},\quad c_3=\frac{1}{16}.
\end{equation}

\end{document}